\documentclass[smallextended]{svjour3}
\usepackage{graphicx}

\title{Improving all-reduce collective operations for imbalanced process arrival patterns}
\author{Jerzy Proficz}
\institute{Jerzy Proficz
\at Gdansk University of Technology\\
Centre of Informatics - Tricity Academic Supercomputer \& networK (CI TASK)\\
11/12 Gabriela Narutowicza Street, 80-233 Gda{\'{n}}sk, Poland\\
\email{j.proficz@task.gda.pl}, Tel.: +48 58 3486343}

\begin{document}
\maketitle
\begin{abstract}
Two new algorithms for the all-reduce operation, optimized for imbalanced process arrival patterns (PAPs) are presented: (i) sorted linear tree (SLT), (ii) pre-reduced ring (PRR) as well as a new way of on-line PAP detection, including process arrival time (PAT) estimations and their distribution between cooperating processes was introduced. The idea, pseudo-code, implementation details, benchmark for performance evaluation and a real case example for machine learning are provided. The results of the experiments were described and analyzed, showing that the proposed solution has high scalability and improved performance in comparison with the usually used ring and Rabenseifner algorithms.
\end{abstract}
\keywords{All-reduce, Pre-reduced ring, Sorted linear tree, Process arrival pattern, MPI}

\section{Introduction}
Collective communication~\cite{MPICollective} is frequently used by the programmers and designers of parallel programs, especially in high performance computing (HPC) applications related to scientific simulations and data analysis, including machine learning calculations. Usually, collective operations, e.g. implemented in MPI~\cite{MPI}, are based on algorithms optimized for the simultaneous entering of all participants into the operation, i.e. they do not take into consideration possible differences in process arrival times (PATs), thus, in real environment, where such imbalances are ubiquitous, they can have significant performance issues. It is worth to note that well performing algorithms for the balanced times, work poorly in the opposite case~\cite{Faraj2008}.

\begin{figure}[!h]
\includegraphics[width=12cm]{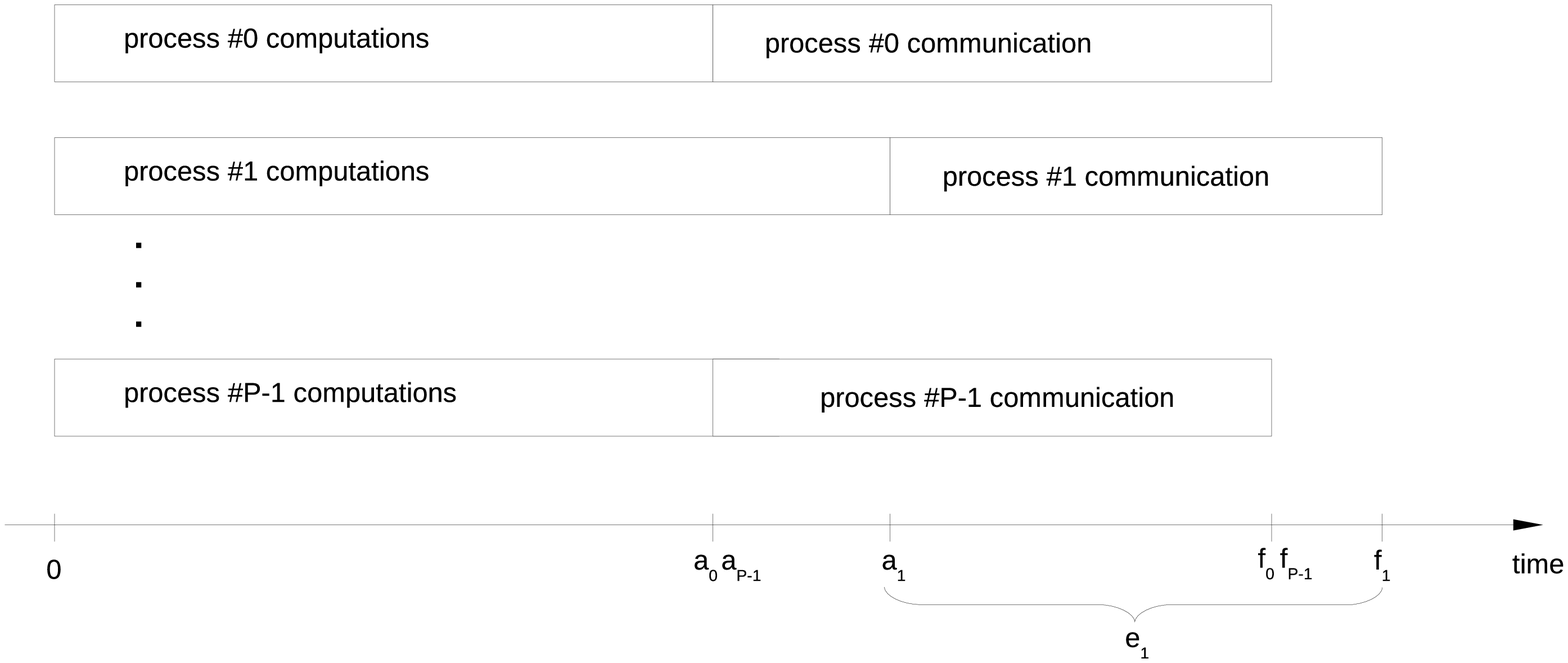} 
\caption{Example of a process arrival pattern with the process \#1 delayed, where $P$ is the number of cooperating processes, $a_i$, $f_i$ and $e_i$ are respectively arrival, finish and elapsed times of process $i$ for the performed collective communication operation}
\label{fig:pap-ex}
\end{figure}

Fig.~\ref{fig:pap-ex} presents an example of a typical execution of a distributed program, where after the computation phase all processes exchange data with each other using some kind of collective communication operation, e.g. all-reduce. We can observe that even for the same computation volume, different processes arrive at the communication phase in different time, in the example processes $1$ is the slowest. Sometimes such differences can be observed in the communication, where the data exchange time is shorter for process 1 than for the other processes.

As a contribution of this paper, we present two new algorithms for the all-reduce operation, optimized for imbalanced process arrival patterns (PAPs): sorted linear tree (SLT) and pre-reduced ring (PRR). We described their idea, pseudo-code, implementation details, benchmark for their evaluation as well as a real case example related to machine learning. Additionally we introduced a new way of on-line PAP detection, including PAT estimations and their distribution among cooperating processes.

The following section presents the related works in the subject, the next one describes the used computation and communication model, section~\ref{sec:algs} presents the proposed algorithms, section~\ref{sec:experiments} provides the evaluation of the algorithms using a benchmark, section~\ref{sec:ml} shows a real case example of the algorithms' utilization, and the last section presents the final remarks.

\section{Related works}
We grouped the related works into three areas: the all-reduce operation in general, i.e. the review of the currently used algorithms in different implementations of MPI~\cite{MPI}, then we describe the current state of the art in process arrival patterns (PAPs), and finally we present the works related to process arrival times (PATs) on-line monitoring and estimation.

\subsection{All-reduce operation}
All-reduce operation is one of the most common collective operations used in HPC software~\cite{Faraj2008}. We can define it as a reduction in a vector of numbers using a defined operation, e.g. sum, which needs to be cumulative, and then distribution of the result into all participating processes, in short: all-reduce = reduce + broadcast. There are plenty of all-reduce algorithms, Table~\ref{tab:ar-mpi} summarizes the ones used in two currently most popular, open source MPI implementations: OpenMPI~\cite{OpenMPI} and MPICH~\cite{MPICH}.

\begin{table}[h]
\caption{All-reduce algorithms implemented in OpenMPI~\cite{OpenMPI} and MPICH~\cite{MPICH}}
\label{tab:ar-mpi}
\begin{tabular}{|p{3.2cm}|p{3.8cm}|p{3.8cm}|}
\hline
Selection criteria & OpenMPI & MPICH\\\hline
Short messages & Recursive doubling~\cite{Thakur2005} & Recursive doubling~\cite{Thakur2005}\\\hline
Long messages & Ring~\cite{Thakur2005} or Segmented ring~\cite{OpenMPI} & Rabenseifner~\cite{Rabenseifner2004}\\\hline
Non-commutive reduce operation & Nonoverlapping~\cite{OpenMPI} & Recursive doubling~\cite{Thakur2005}\\\hline
Others & Basic linear~\cite{OpenMPI}, Segmented linear tree~\cite{OpenMPI} & {} \\\hline
\end{tabular}
\end{table}

\begin{figure}[!h]\sidecaption
\includegraphics[width=8cm]{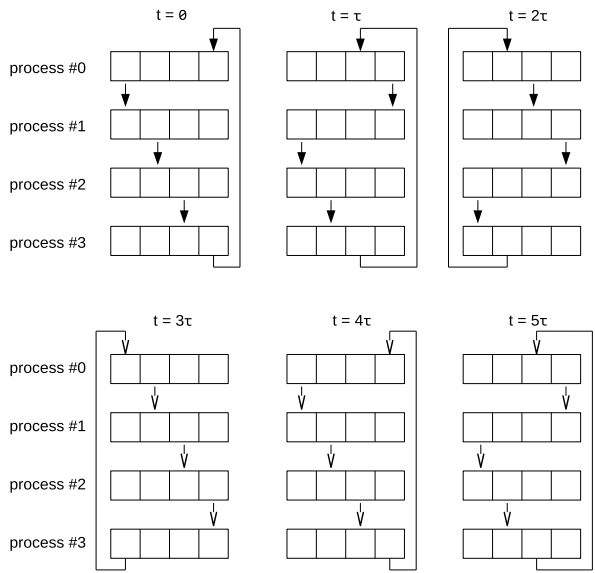} 
\caption{Example of a ring algorithm execution, showing data distribution between the processes (one square corresponds to one data segment), a solid arrow represents send and reduce a data segment, an empty arrow represents send and override a data segment, and $\tau$ is the time of transfer and reduction/overriding of a data segment}
\label{fig:ring}
\end{figure}

The \emph{basic linear} algorithm performs linear reduce (flat tree: the root process gathers and reduces all data from the other processes) followed by the broadcast without any message segmentation~\cite{OpenMPI}. \emph{Segmented linear tree} creates pipeline between the participating processes, where the data are split into segments and sent from process $0$ to $1$ to $2$~\ldots to $P-1$. The \emph{nonoverlapping} algorithm uses default reduce followed by the broadcast, these operations are not overlapping: the broadcast is performed sequentially after the reduce, even if both use segmentation and some segments are ready after the reduce~\cite{OpenMPI}. \emph{Recursive doubling} (a.k.a. \emph{butterfly} and \emph{binary split}) is performed in rounds, in every round each process exchanges and reduces its data with another, corresponding process, whose rank changes round by round, after $log_2P$ rounds ($P$ is the total number of participating processes) all data are reduced and distributed to the processes~\cite{Thakur2005}. \emph{Ring} is also performed in rounds, the data are divided into $P$ segments and each process sends one segment per round (the segment index depending on the round number and the process rank) to the next process (the processes are usually ordered according to their ranks). At the beginning, there are performed $P-1$ rounds, where a segment delivery is followed by its data reduction, thus after these rounds, each process has exactly one segment fully reduced, and needs to forward it to other processes. This is performed during next $P-1$ rounds, where the data are gathered, i.e. the processes transfer the reduced segments, thus the final result is delivered to all processes~\cite{Thakur2005}. Fig.~\ref{fig:ring} presents an example of a ring execution. \emph{Segmented ring} is similar to the ring, but after the segmentation related to process number, an additional one is performed for pipeline effect between corresponding processes~\cite{OpenMPI}. In the \emph{Rabenseifner} algorithm the reduction is performed in two phases, first the scatter and reduce of data are executed (using recursive doubling on divided data) and then the data gathering takes place (using recursive halving on gathered data), where the reduced data come back to all processes~\cite{Rabenseifner2004}.

\subsection{Imbalanced process arrival times}
Not much work has been done for imbalanced process arrival patterns analysis. To the best of our knowledge, the following four papers cover the research performed in this area.

In~\cite{Faraj2008} Faraj et al. performed advanced analysis of \emph{process arrival patterns} (PAPs) observed during execution of typical MPI programs. They executed a set of tests proving that the differences between \emph{process arrival times} (PATs) in operations of the collective communication are significantly high and they influence the performance of the underlying computations. The authors defined a PAP to be imbalanced for a given collective operation with a specific message length when its imbalance factor (a ratio between the highest difference between the arrival times of the processes and time of the simple (point-to-point) message delivery between each other) is larger than 1. The authors provided examples of typical HPC benchmarks, e.g. NAS, LAMMPS or NBODY, where imbalance factor, during their execution in a typical cluster environment, equals 100 or even more. They observed that, such behavior usually cannot be controlled directly by the programmer, and the imbalances are going to occur in any typical HPC environment. The authors proposed a mini-benchmark for testing various collective operations and found out the conclusion that the algorithms which perform better with balanced PAPs tend to behave worse when dealing with imbalanced ones. Finally, they proposed solution: their self-tuning framework -- STAR-MPI~\cite{Faraj2006}, which includes a large set of various implementations of collective operation and can be used for different PAPs, with automatic detection of the most suitable algorithm. The framework efficiency was proved by an example of tuned all-to-all operations, where the performance of the set of MPI benchmarks was significantly increased.

As a continuation of the above work Patarasuk et al. proposed a new solution for broadcast operation used in MPI application concerning imbalanced PAPs of the cooperating processes~\cite{Patarasuk2008}. The authors proposed a new metric for the algorithm performance: competitive ratio -- $PERF(r)$, which describes the influence of the imbalanced PATs for the algorithm execution, regarding the behavior for its worst case PAT. They evaluated well known broadcast algorithms, using the above metric, and presented two new algorithms, which have constant (limited) value of the metric. The algorithms are meant for large messages, and use the subsets of cooperating processes to accelerate the overall process: the data are sent to the earliest processes first. One of the algorithms is dedicated for non-blocking (\emph{arrival\_nb}) and other for blocking (\emph{arrival\_b}) message passing systems. The authors proposed a benchmark for algorithms’ evaluation, which introduces random PAPs and measures their impact on the algorithm performance. The experiments were performed using two different 16-node compute clusters (one with Infiniband and other with Ethernet interconnecting network), and 5 broadcast algorithms, i.e. \emph{arrival\_b}, flat, linear, binomial trees and the one native to the machine. The results of the experiments showed the advantage of the \emph{arrival\_b} algorithm for large messages and imbalanced PAPs.

In~\cite{Marendic2012} Marendic et al. focused on an analysis of reduce algorithms working with imbalanced PATs. They assumed atomicity of reduced data (the data cannot be split into segments and reduced piece by piece), as well as the Hockney~\cite{Hockney1994} model of message passing (time of message transmission depends on the link bandwidth: $\beta$ and constant latency: $\alpha$, with an additional computation speed parameter: $\gamma$) and presented related works for typical reduction algorithms. They proposed a new static load balancing optimized reduction algorithm requiring a priori information about current PATs of all cooperating processes. The authors performed a theoretical analysis proving the algorithm is nearly optimal for the assumed model. They showed that the algorithm gives the minimal completion time under the assumption that the corresponding point-to-point operations start exactly at the same time for any two communicating processes. However if the model introduces a delay of the receive operation in comparison with the send one, which seems to be the case in real systems, the algorithm does not utilize this additional time in receiving process, although, in some cases, it could slightly improve the performance of the overall reduce operation. The other proposed algorithm, presented by the authors: a dynamic load balancing can operate under the limited knowledge about PATs, being able to atomically reconfigure the message passing tree structure while performing reduce operation using auxiliary short messages for signaling the PATs between the cooperating processes. The overhead is minimal in comparison with the gains of the PAP optimization. Finally a mini-benchmark was presented and some typical PAPs were examined, the results showed the advantage of the proposed dynamic load balancing algorithm versus other algorithms: binary tree and all-to-all reduce.

In~\cite{Marendic2016} Marendic et al. continued the work with optimization of the MPI reduction operations dealing with the imbalanced PAPs. The main contribution is a new algorithm, called Clairvoyant, scheduling the exchange of data segments (fixed parts of reduced data) between reducing processes, without assumption of data atomicity, and taking into account PATs, thus causing as many as possible segments to be reduced by the early arriving processes. The idea of the algorithm bases on the assumption that the PAP is known during process scheduling. The paper provided a theoretical background for the PAPs, with its own definition of the time imbalances, including a PAT vector, absolute imbalance, absorption time as well as their normalized versions, followed by the analysis of the proposed algorithm, and its comparison to other typically used reduction algorithms. Its pseudo-code was described and the implementation details were roughly provided with two examples of its execution for balanced and imbalanced PAPs. Afterwards the performed experiments were described, including details about used mini-benchmark and the results of practical comparison with other solutions (typical algorithms with no support for imbalanced PAPs) were provided. Finally, the results of the experiments showing advantage of the proposed algorithm were presented and discussed. 

\subsection{Process arrival time estimation}
The PAP collective communication algorithms require some knowledge about the PATs for their execution. Table~\ref{tab:pat-detection} presents the summary of the approaches used in the works described above.

\begin{table}[h]
\caption{Approaches for PAP detection}
\label{tab:pat-detection}
\begin{tabular}{|p{4cm}|p{7cm}|}
\hline
Operation & PAP detector\\\hline
All-to-all~\cite{Faraj2008} & Uses STAR-MPI overall efficiency indicators\\\hline
Broadcast~\cite{Patarasuk2008} & Uses its own messaging for PAT signaling\\\hline
Local redirect~\cite{Marendic2012} & Uses its own messaging for PAT signaling\\\hline
Clairvoyant~\cite{Marendic2016} & None, suggested usage of the static analysis or SMA\\\hline
\end{tabular}
\end{table}

In~\cite{Faraj2008} (dedicated for an all-to-all collective operation), there is assumption about the call site (a place in the code where the MPI collective operation is called) paired with the message size that they have a similar PAPs for the whole program execution, or at least their behavior changes infrequently. The proposed STAR-MPI~\cite{Faraj2006} system periodically assesses the call site performance (exchanging measured times between processes) and adapts a proper algorithm, trying one after another. The authors claim that it requires 50--200 calls to provide desired optimization. This is a general approach and it can be used even for other performance issues, e.g. network structure adaptation.

In~\cite{Patarasuk2008} (dedicated for a broadcast operation), the algorithm uses additional, short messages sent to root process signaling process readiness for the operations. In case the some processes are ready, the root performs sub-group broadcast, thus the a priori PATs are not necessary for this approach. Similar idea is used in~\cite{Marendic2012} (dedicated for a reduce operation) where the additional messages are used not only to indicate readiness, but also to redirect delayed processes.

In~\cite{Marendic2016} (dedicated for a reduce operation), the algorithm itself does not include any solution for the PAT estimation, that is why it is called Clairvoyant, but the authors assume recurring PAPs and give the suggestion that there can be used simple moving averages (SMA) approximation. This solution requires the additional communication to exchange the PAT values, what is performed every $k$ iterations, thus introducing the additional communication time. The authors claim that the speedup introduced by the usage of the algorithm overcomes this cost, and provide some experimental results showing the total time reduction in the overall computations.

\section{Computation and communication model}\label{sec:model}
We assume usage of the message-passing paradigm, within a homogeneous environment, where each communicating process is placed on a separated compute node. The nodes are connected by a homogeneous communication network. Every process can handle one or more threads of control communicating and synchronizing with each other using shared memory mechanisms. However there is no shared memory accessible simultaneously by different processes.

As a \emph{process arrival pattern} (PAP) we understand the timing of different processes arrivals for a concrete collective operation, e.g. all-reduce in an MPI program. We can evaluate a given PAP by measuring the \emph{process arrival time} (PAT) for each process. Formally, a PAP is defined as the tuple $(a_0, a_1, \ldots a_{P-1})$, where $a_i$ is a measured PAT for process $i$, while $P$ is the number of processes participating in the collective operation. Similarly \emph{process exit pattern} (PEP) is defined as the tuple $(f_0, f_1, \ldots f_{P-1})$, where $f_i$ is the time when process $i$ finishes the operation~\cite{Faraj2008}. Fig.~\ref{fig:pap-ex} presents an example of arrival and exit patterns.

For each process participating in a particular operation Faraj et al. define the \emph{elapsed time} as $e_i=f_i-a_i$, and the \emph{average elapsed time} for the whole PAP: $\bar{e}=\frac{1}{P}\sum_{i=0}^{P-1}e_i$~\cite{Faraj2008}. This is a mean value of time spent for communication by each process, the rest of the time is used for computations. Thus minimizing the elapsed times of the participating processes decreases the total time of program execution and, in our case, is the goal of the optimization.

For an all-reduce operation, assuming $\delta$ to be the time of sending the reduced data between any two processes and only one arbitrary chosen process $k$ is delayed (others have the same arrival time, see an example in Fig.~\ref{fig:pap-ex}, where $k=1$), we can estimate the lower bound of $\bar{e}$ as $\bar{e}_{lo}=a_k-a_o+\delta$, where $a_o$ is the time of arrival of all processes except $k$. On the other hand, assuming $\Delta$ to be time of an all-reduce operation for perfectly balanced PAT ($a_i=a_j$, for all $i,j\in\langle 0, k-1\rangle$), we can estimate the upper bound of $\bar{e}$ as $\bar{e}_{up}=a_k-a_o+\Delta$. Thus using PAP optimized all-reduce algorithm can decrease the elapsed time by $\Delta-\delta$ or less. E.g. for a typical ring all-reduce algorithm, working on 16~processes, 4\,MB data size and 1\,Gbps Ethernet network we can measure $\Delta=$45.7\,ms and $\delta=$18.2\,ms, thus using PAP optimized algorithm can save at most 27.5\,ms of average elapsed time, no matter how slow the delayed process is.

Furthermore, we assume a typical iterative processing model with two phases: the computation phase where every process performs independent calculations, and the communication phase where the processes exchange the results, in our case, using all-reduce collective operation. These two phases are combined into an iteration, which is repeated sequentially during the processing. We assume that the whole program execution consists of $N$ iterations.

Normally, during a computation phase, the message communication between processes/nodes is suspended. Nevertheless, each process can contain many threads carrying on the parallel computations exploiting shared memory for synchronization and data exchange. Thus, during this phase the communication network connecting nodes is unused and can be utilized for exchange of additional messages, containing information about a progress of computations or other useful events (e.g. a failure detection).

\begin{figure}[!h]
\includegraphics[width=12cm]{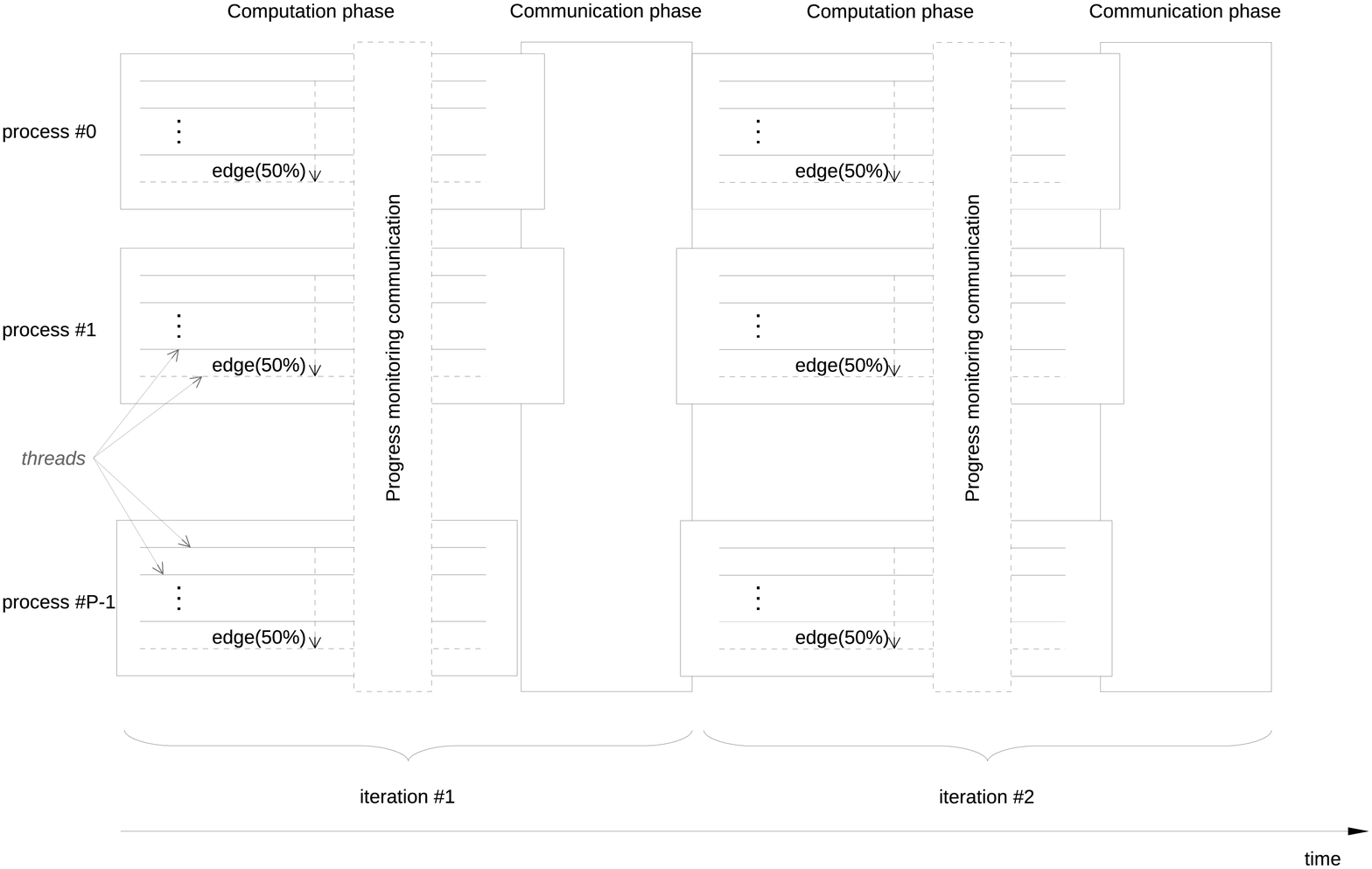} 
\caption{Iterative processing model (solid lines), extended by progress monitoring (dashed lines)}
\label{fig:processing-model}
\end{figure}

Thus, we introduce an additional thread which is responsible for inter-node communication during computation phase. It monitors the progress of the phase on its node, estimates the remaining computation time and exchanges this knowledge with other processes on the cooperating nodes. Gathering this information, every thread can approximate a process arrival pattern (PAP) for itself and other processes (see Fig.~\ref{fig:processing-model}).

For monitoring purposes the computation threads need to pass the status of current iteration processing calling a special function: edge(). We assume that for all processes this call is made after a defined part of performed computations e.g. 50\%, while the exact value is passed as the function parameter. In our implementation the edge() function is executed in some kind of callback function, reporting a status of the iteration progress.

Beside the PAP monitoring and estimation functions the additional thread can be used for other purposes. In our case, the proposed all-reduce algorithms, described in section~\ref{sec:algs}, are working much better (performing faster message exchange) when connections between the communicating processes/nodes are already established, thus we introduced an additional warm-up procedure where a series of messages are transferred in the foreseen directions of communication. This operation is performed by the thread after the exchange of the progress data.

\begin{figure}[!h]\sidecaption
\includegraphics[width=8cm]{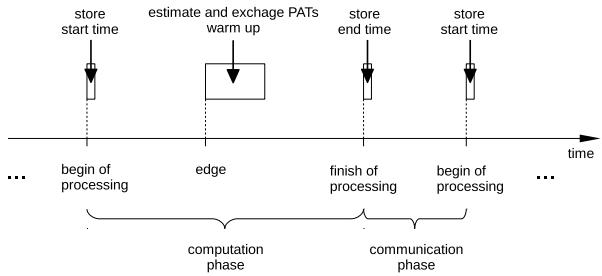} 
\caption{Events and actions performed in the additional thread}
\label{fig:add-thread}
\end{figure}

Thus the analysis of the above requirements implemented in the additional thread, shows that the thread should react accordingly to the following events (see Fig.~\ref{fig:add-thread}):
\begin{itemize}
\item the beginning of processing, when the thread is informed about the computation phase start and when it stores the timestamp for the further time estimation,
\item the edge, in the middle of processing, when the thread estimates the ongoing computation phase time for its process, exchanges this information with other processes and performs the warm up procedure for establishing connections to speed up the message transfer in the coming collective operation,
\item the finish of processing, when the thread is informed about the end of the computation phase and when it stores the timestamp for the further time estimations.
\end{itemize}

\section{New all-reduce algorithms optimized for PAPs}\label{sec:algs}
In this section we introduce two new algorithms for all-reduce operations, optimized for a PAP observed during the computation phase: (i) sorted linear tree (SLT), (ii) pre-reduced ring (PRR). Both of them are based on the well known, and widely used regular all-reduce algorithms: linear tree~\cite{OpenMPI} and ring~\cite{Thakur2005}, respectively, and have similar communication and time complexity.

\subsection{Sorted linear tree (SLT)}
The algorithm is an extension of the linear tree~\cite{OpenMPI}, which transfers the data segments sequentially through processes exploiting the pipeline parallel computation model. The proposed modification causes the processes to be sorted by their arrival times. While the faster processes start the communication earlier, the later ones have more time to finish the computations.

Fig.~\ref{fig:slt} presents pseudo-code of the algorithm. At the beginning new identifiers are assigned to the processes according to the arrival order, then the data are split into equal segments, in our case we assume $P$ segments, where $P$ is a number of the cooperating processes. Afterwards the reduce loop is started (lines: 1-6), where the data segments are transferred and reduced, and then the override loop is executed (lines: 7-12), where the segments are distributed back to all processes.

\begin{figure}[!h]
\fbox{\parbox{.97\textwidth}{\sffamily
input parameters:\\
$P$ -- number of processes or nodes, we assume one process per node\\
$a_i$ -- arrival time of process $i$\\
$d_x$ -- $x$-th of $P$ data segments to be reduced, e.g. $d_0$ - first data segment\\
$id$ -- a new process rank after sorting according to arrival times\\

variables:\\
$si$ -- segment index\\

// reduce loop\\
1.~~for $i:=0$ to $P-1$\\
2.~~~~$si := i\ mod\ P$\\
3.~~~~if $id \neq 0$ then // not the first process\\
4.~~~~~~$rd$ = receive\_segment() // blocking\\
5.~~~~~~reduce($d_{si}$, $rd$)\\
6.~~~~send\_segment($d_{si}$, $(id+1)\ mod\ P$) // non-blocking\\

// override loop\\
7.~~for $i:=0$ to $P-1$\\
8.~~~~$si := i\ mod\ P$\\
9.~~~~if $id \neq P-1$ then // not the last process\\
10.~~~~~$d_{si} :=$ receive\_segment() // blocking\\
11.~~~if $id < P-2$ then // not the last or penultimate process\\
12.~~~~~send\_segment($d_{si}$, $(id+1)\ mod\ P$) // non-blocking
}}
\caption{Pseudo-code of the sorted linear tree (SLT) algorithm}
\label{fig:slt}
\end{figure}

\begin{figure}[!h]
\includegraphics[width=12cm]{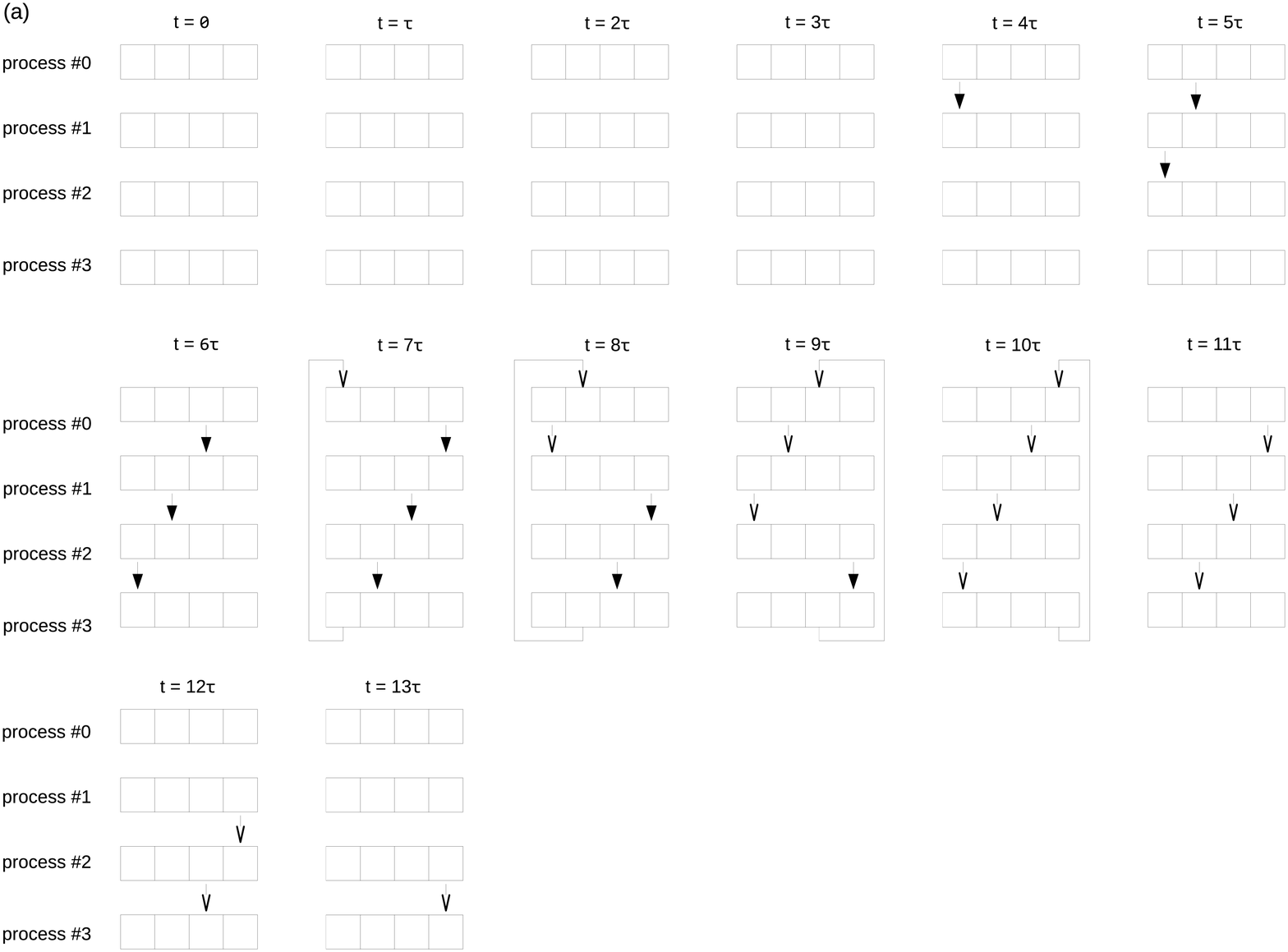} 
~\\
\includegraphics[width=12cm]{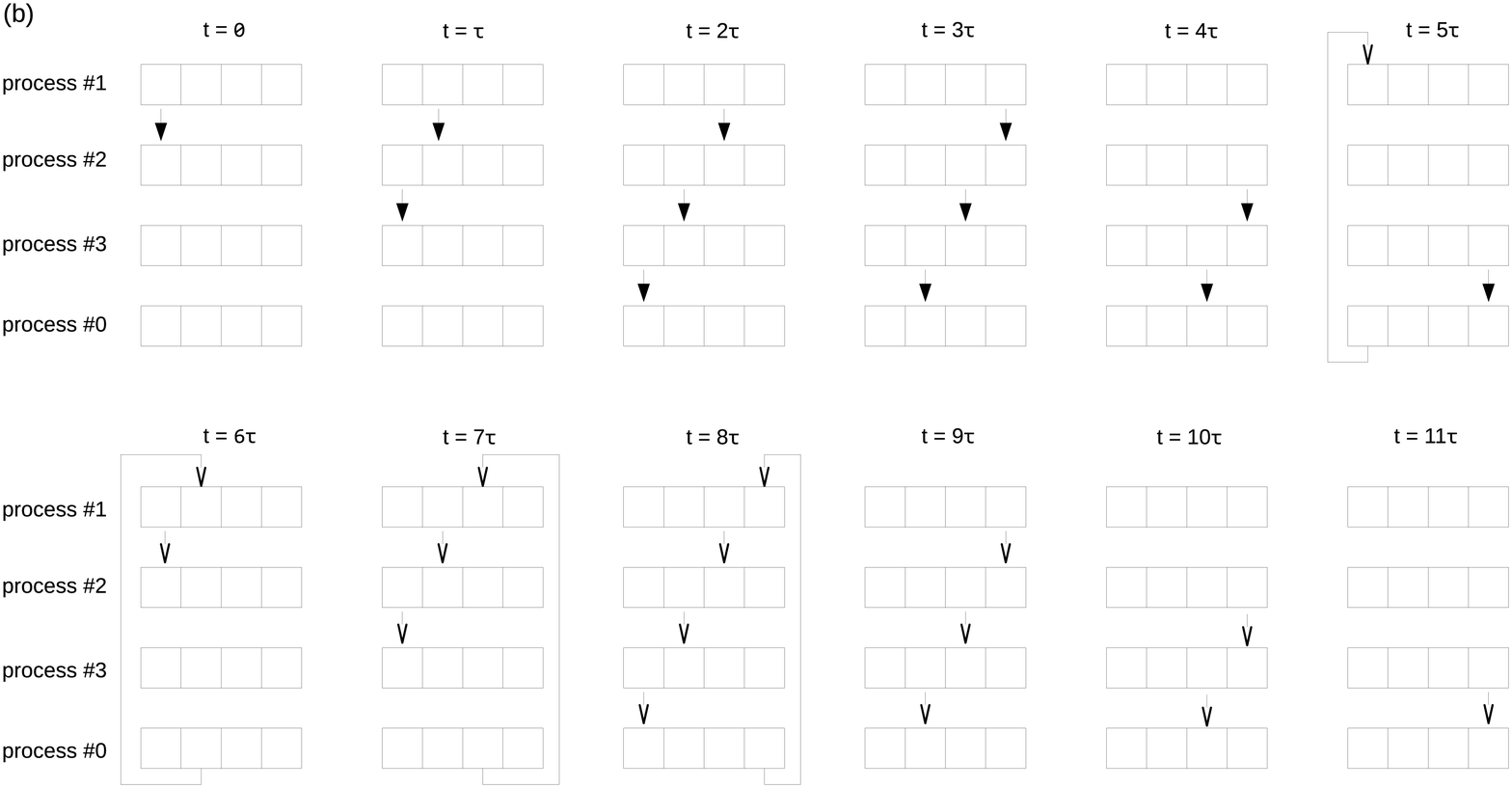} 
\caption{Example of (a) linear tree and (b) sorted linear tree (SLT) algorithm executions, showing data distribution between the processes (one square corresponds to one data segment), where process \#0 is delayed by $4\tau$, a solid arrow represents send and reduce a data segment, an empty arrow represents send and override a data segment, and $\tau$ is the time of transfer and reduction/overriding of a data segment}
\label{fig:sltex}
\end{figure}

Let's assume $\tau$ is a time period required for transfer and reduction/overriding of one data segment between any two processes. Fig.~\ref{fig:sltex}{}b presents an example of SLT execution, where the process $0$ arrival time is delayed for $4\tau$ in comparison with all other processes. Using the regular linear tree reduction, the total time of the execution would be $14\tau$ (see Fig.~\ref{fig:sltex}{}a), while the knowledge of the PAP and procedure of sorting the processes by their arrival times make the delayed process to be the last one in the pipeline and cause the whole operation time to be decreased to $12\tau$.

\subsection{Pre-reduced ring (PRR)}
This algorithm is an extended version of ring~\cite{Thakur2005}, where each data segment is reduced and then passed to other processes in synchronous manner. The idea of the algorithm is to perform a number of so-called, reducing pre-steps, between faster processes (with lower arrival times), and then the regular processing, like in the typical ring algorithm, is performed.

Fig.~\ref{fig:prr} presents pseudo-code of the algorithm. First of all, the processes are sorted by their arrival times and the new ids are assigned, i.e. initially the processes perform communication in direction from the earliest to the latest ones. Then data are split into $P$ equal segments and for each such a segment the number of pre-steps is calculated (lines: 1-6), its value depending directly on the estimated process arrival times (PATs): $a_i$ and the time of transfer and reduction/overriding of one segment: $\tau$. Knowing the above, the algorithm resolves where each segment starts and finishes its processing, thus two process ids are assigned for each segment: $sp_i$ and $rp_i$, respectively (lines: 7-13).

\begin{figure}[!h]
\fbox{\parbox{.97\textwidth}{\sffamily
input parameters:\\
$\tau$ -- time of transfer and reduction/overriding of one segment\\
$P$ -- number of processes or nodes, we assume one process per node\\
$a_i$ -- arrival time of process $i$\\
$d_x$ -- $x$-th of $P$ data segments to be reduced, e.g. $d_0$ - first data segment\\
$id$ -- a new process rank after sorting according to arrival times\\

variables:\\
$k_i$ -- number of pre-steps in process $i$\\
$si$ -- segment index\\
$sp_j$ -- id of the first process where the slice $j$ is sent\\
$rp_j$ -- process id where the slice j is reduced last time\\

// calculate a number of pre-steps for all processes\\
1.~~$k_{P-1} := 0$\\
2.~~for $i:=P-2$ to 0\\
3.~~~~if $a_{P-1}-a_{i+1} \ge (k_{i+1}+1)\times\tau$ then\\
4.~~~~~~$k_i:=k_{i+1}+1$\\
5.~~~~else\\
6.~~~~~~$k_i:=k_{i+1}$\\

// calculate ids of the processes where a slice id send first time\\
// and ids of the processes where a slice is reduced last time\\
7.~~$i := 0$\\
8.~~for $j:=0$ to $P-1$\\
9.~~~~if $i+k_i \ge j$ then\\
10.~~~~~$sp_j := i$\\
11.~~~else\\
12.~~~~~$i := i+1$\\
13.~~~$rp_j := (sp_j+P-1)\ mod\ P$\\

// calculate what segment start for\\
14. $si := id+k_{id}$\\

// reduce loop\\
15.~for $i:=0$ to $P-1$\\
16.~~~if $sp_{si} \neq id$ then\\
17.~~~~~$rd :=$ receive\_segment() // blocking\\
18.~~~~~reduce($d_{si}$, $rd$)\\
19.~~~if $(rpsi+P-1)\ mod\ P \neq is$ then\\
20.~~~~~send\_segment($d_{si}$, $(id+1)\ mod\ P$) // non-blocking\\
21.~~~$si := (si+P-1)\ mod\ P$\\

// override loop\\
22.~for $i:=0$ to $P-1$\\
23.~~~if $rp_{si} \neq id$ then\\
24.~~~~~$d_{si} :=$ receive\_segment() // blocking\\
25.~~~if $(rp_{si}+P-1)\ mod\ P \neq id$ then\\
26.~~~~~send\_segment($d_{si}$, $(id+1)\ mod\ P$) // non-blocking\\
27.~~~$si := (si+P-1)\ mod\ P$
}}
\caption{Pseudo-code of the pre-reduced ring (PRR) algorithm}
\label{fig:prr}
\end{figure}

Afterwards, initial value of the segment index: $s_i$ is calculated from which every process starts processing. In the case of the regular ring algorithm, its value depends on the process $id$ only, however for PRR it also regards the pre-steps performed by the involved processes (line: 14). Then the reduce loop is started, the pre-steps and the regular steps are performed in one block, where the variable $s_i$ controls which segments are sent, received and reduced (lines: 15-21). Similarly, the override loop is performed, being controlled by the same variable (lines: 22-27).

In the regular ring every process performs $2P-2$ receive and send operations, thus total number is $P\times(2P-2)$. In case of PRR, this total number is the same, however the faster processes (the ones which finished computation earlier) tend to perform more communication while executing the pre-steps. In the case, when the arrival of the last process is largely delayed to the next one (i.e. more than $P\times\tau$) it performs only $P-1$ send and receive operations, while every other process does $2P-1$ ones.

\begin{figure}[!h]
\includegraphics[width=9.5cm]{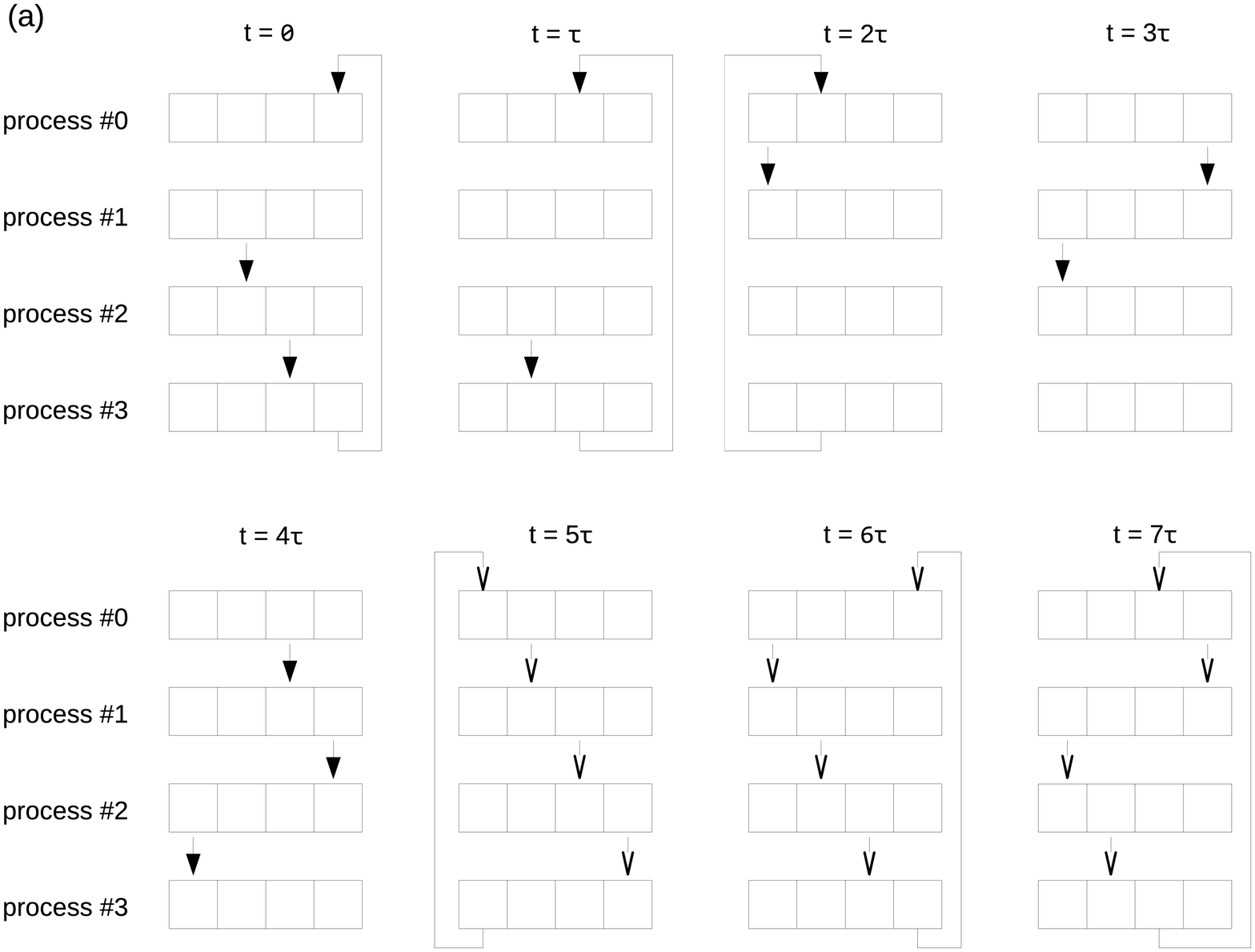} 
~\\
\includegraphics[width=9.5cm]{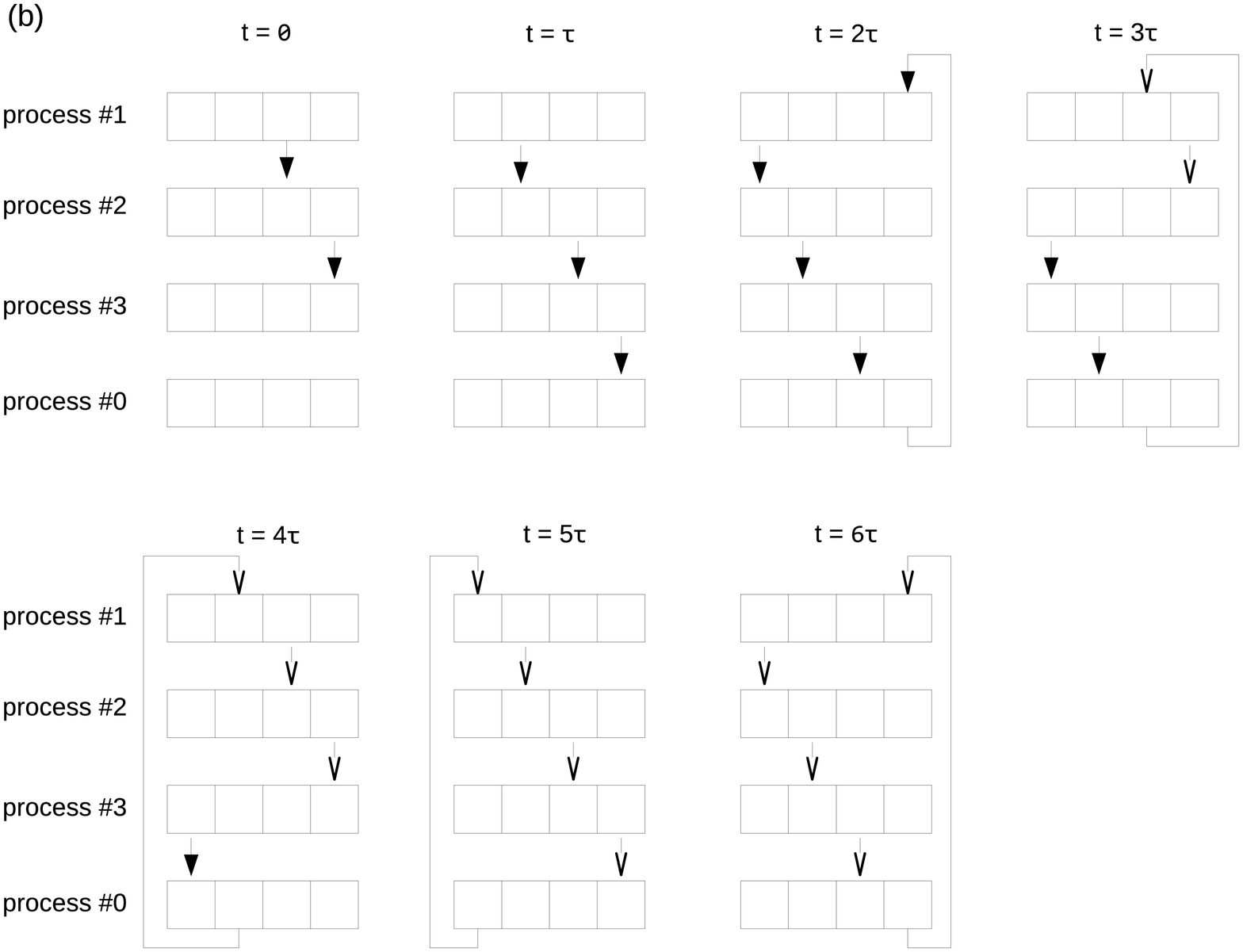} 
\caption{Example of (a) ring and (b) pre-reduced ring (PRR) algorithm executions, showing data distribution between the processes (one square corresponds to one data segment), where process \#0 is delayed by $2\tau$, a solid arrow represents send and reduce a data segment, an empty arrow represents send and override a data segment, and $\tau$ is the time of transfer and reduction/overriding of a data segment}
\label{fig:prrex}
\end{figure}

Fig.~\ref{fig:prrex}{}b presents an example of PRR execution, where the process $0$ arrival time is delayed for $2\tau$ ($2\times$ time of transfer and reduction/overriding of one data segment) in comparison with all other processes. Using the regular ring reduction, the total time of the execution would be $8\tau$ (see Fig.~\ref{fig:prrex}b), while the knowledge of the PAP, performing two additional reduction pre-steps and setting the delayed process to be the last one in the pipeline, causes the whole operation time to be reduced to $7\tau$.

\section{The experiments}\label{sec:experiments}
The above algorithms were implemented and tested in a real HPC environment, the following subsections describe a proposed benchmark (including its pseudo-code and implementation details), the experiment's setup and provide the discussion about the observed results.

\subsection{The benchmark}
Fig.~\ref{fig:bench} presents pseudo-code of the proposed benchmark evaluating the performance of the proposed algorithms in the real HPC environment using MPI~\cite{MPI} for communication purposes. For every execution there is a sequence of repeated iterations consisting of the following actions:
\begin{description}
\item[line 2:] data generation, where the data are randomly assigned,
\item[line 3:] calculation of the emulated computation time, ensuring that the progress monitoring communication (the additional 100\,ms) is completely covered by the computation phase,
\item[lines 4--8:] the delay mode is applied, in `one-late' mode only one process (id: 1) is delayed for $maxDelay$, while in `rand-late' mode all processes are delayed for random time up to $maxDelay$,
\item[lines 9--10:] two MPI\_Barrier() calls, making sure all the processes are synchronized,
\item[lines 11--13:] the emulation of computation phase by using sleep() (usleep() in the implementation) function, including call to edge() function (see section~\ref{sec:model}), providing the progress status to the underlying monitoring thread,
\item[lines 14--16:] call to the all-reduce algorithm implementation, including the commands for the time measurements, for the benchmark purposes we assumed sum as the reduce operator,
\item[line 17:] checking the correctness of the performed all-reduce operation, using regular MPI\_Allreduce() function,
\item[lines 18--19:] calculation of the elapsed time for the current and for all processes using MPI\_Allgather() function,
\item[line 20:] saving the average elapsed time into the $results$ vector.
\end{description}

\begin{figure}[!h]
\fbox{\parbox{.97\textwidth}{\sffamily
input parameters:\\
$size$ -- number of elements (floats) in reduced data\\
$N$ -- number of iterations\\
$mode$ -- mode of delay: one-late/rand-late\\
$maxDelay$ -- maximal delay of the process(es)\\
$algorithm$ -- tested algorithm, one of ring, Rabenseifner, PRR, SLT\\
$P$ -- number of processes\\
$id$ -- process id -- MPI rank: $0$\dots$P-1$\\

output:\\
$results$ -- vector of measured average elapsed times for the given algorithm\\

variables:\\
$halfTime$ -- 50\% of the emulated computation time\\
$startTime$ -- start time of measurement\\
$endTime$ -- end time of measurement\\
$myElapsedTime$ -- elapsed time measured in the current process\\
$allElapsedTimes$ -- vector of the measured times of all processes\\
$data$ -- vector of data to be reduced\\

1.~~for $i := 1$ to $N$\\
2.~~~~$data$ := generateRandomData($size$)\\
3.~~~~$halfTime = (maxDelay+100\,ms)/2$\\
4.~~~~if $mode =$ "one-late" then\\
5.~~~~~~if $id = 1$ then\\
6.~~~~~~~~$halfTime := halfTime+maxDelay$\\
7.~~~~else // $mode =$ "rand-late" \\
8.~~~~~~$halfTime := halfTime +$ random($0$\dots$maxDelay$)$/2$\\
9.~~~~MPI\_Barrier()\\
10.~~~MPI\_Barrier()\\
11.~~~sleep($halfTime$)\\
12.~~~edge($0.5$)\\
13.~~~sleep($halfTime$)\\
14.~~~$startTime :=$ MPI\_Wtime()\\
15.~~~makeAllReduce($algorithm$, $data$)\\
16.~~~$endTime :=$ MPI\_Wtime()\\
17.~~~checkCorrectness($data$)\\
18.~~~$myElapsedTime := endTime-startTime$\\
19.~~~MPI\_Allgather($allElapsedTimes$, $myElapsedTime$, \ldots)\\
20.~~~results[i] := average($allElapsedTimes$)
}}
\caption{Pseudo-code of the performance benchmark}
\label{fig:bench}
\end{figure}

For the comparison purposes we used two typical all-reduce algorithms: ring~\cite{Thakur2005} and Rabenseifner~\cite{Rabenseifner2004}, they are implemented in the two most popular open source MPI implementations: OpenMPI~\cite{OpenMPI} and MPICH~\cite{MPICH} respectively, and used for large input data size. To the best knowledge of the author there are no PAP optimized all-reduce algorithms described in the literature.

The benchmark was implemented in C language v.~C99, compiled using GCC v.~7.1.0, with the maximal code optimization (-O3). The program uses OpenMPI v.~2.1.1 for processes/nodes message exchange and POSIX Threads v.~2.12 for intra-node communication and synchronization, moreover for managing of dynamic data structures GLibc v.~2.0 library was used. In this implementation the reduce operation is based on a sum (equivalent of using MPI\_SUM for $op$ parameter in MPI\_Allreduce()).

\subsection{Environment and test setup}\label{sec:bench-setup}
The benchmark was executed in a real HPC environment using cluster supercomputer Tryton, placed in Centre of Informatics - Tricity Academic Supercomputer and NetworK (CI TASK) at Gdansk University of Technology in Poland~\cite{Krawczyk2015}. The cluster consists of homogeneous nodes, where each node contains 2 processors (Intel Xeon Processor E5 v3, 2.3GHz, Haswell), with 12~physical cores (24~logical ones, due to Hyperthreading technology) and 128\,GB RAM memory. In total the supercomputer consists of 40 racks with 1,600~servers (nodes), 3,200~processors, 38,400~compute cores, 48~accelerators and 202\,TB~RAM memory. It uses fast FDR 56\,Gbps InfniBand in fat tree topology and 1\,Gbps Ethernet for networking. Total computing power is 1.48\,PFLOPS. The cluster weighs over 20 metric tons.

The experiments were performed using a subset of the nodes, with HT switched off, grouped in one rack and connected to each other trough a 1\,Gbps Ethernet switch (HP J9728A 2920-48G). The benchmark input parameters were set up to the following values:
\begin{description}
\item[$algorithm$:] ring, Rabenseifner, pre-reduced ring (PRR) and sorted linear tree (SLT);
\item[$size$:] (of data vector) 128\,K, 512\,K, 1\,M, 2\,M, 4\,M, 8\,M of floats (4~bytes long);
\item[$mode$:] (of process delay) one-late (where only one process is delayed by $maxDelay$) and rand-late (where all processes are delayed randomly up to $maxDelay$);
\item[$maxDelay$:] (of processes arrival times) 0, 1, 5, 10, 50, 100, 500, 1000\,ms;
\item[$P$:] (number of processes/nodes) 4, 6, 8, 10, 12, 16, 20, 24, 28, 32, 36, 40, 44, 48;
\item[$N$:] (number of iterations) 64--256, depending on $maxDelay$ (more for lower delay); 
\end{description}

\subsection{Benchmark results}
Table~\ref{tab:alg-cmp} presents the results of the benchmark execution for 1\,M of floats of reduced data, 1\,Gbps Ethernet network, where only one process was delayed on 48~nodes in a cluster environment of Tryton~\cite{Krawczyk2015} HPC computer. The results are presented as absolute values of average elapsed time: $\bar{e}_{alg}$ and speedup: $s_{alg}$, in comparison with ring algorithm $s_{alg}=\frac{\bar{e}_{ring}}{\bar{e}_{alg}}$, where $alg$ is the evaluated algorithm.

\begin{figure}[!h]\sidecaption
\includegraphics[width=8cm]{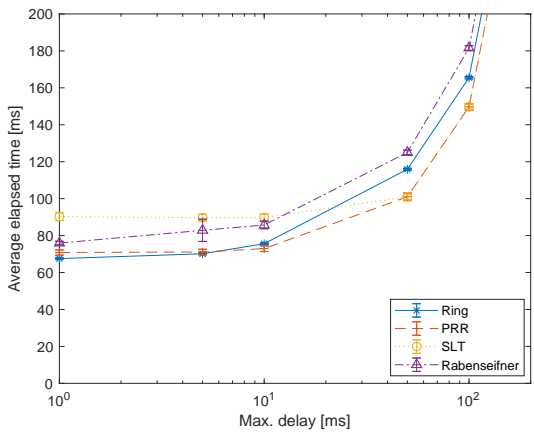} 
\caption{Benchmark results showing influence of the increasing maximum delay on average elapsed times of the tested algorithms. The experiments were performed on 48~nodes for 1\,Gbps Ethernet network, one process delayed, and message size: 1024\,K of float numbers. The error bars are set to $\pm{}2\sigma$ (95\% of the measurements for the normal distribution)}
\label{fig:bench-delay}
\end{figure}

In this setup the ring algorithm seems to be more efficient than the Rabenseifner, thus in further analysis we use the former for reference purposes. For more balanced PAPs, where the delay is below 5\,ms the proposed algorithms perform worse than the ring, it is especially visible for SLT (about 25\% slower), however PRR shows only slight difference (below 5\%). On the other hand, when the PAP is more imbalanced with the delay over 10\,ms, both algorithms perform much better (up to 15\% faster than the ring). For the really high delays, the results stabilize providing about 17\,ms of average elapsed time savings, causing the total speedup to be lower. The comparison of influence of changes in the (maximum) delay on the algorithms' performance is presented in Fig.~\ref{fig:bench-delay}.

\begin{table}[h]
\caption{Benchmark results for 1\,M of floats of reduced data, 1\,Gbps Ethernet network, only one process delayed and 48~processes/nodes. Maximum delay is measured in ms and each result consists of two values: the average elapsed times in ms and speedup in comparison with the ring algorithm ($\frac{\bar{e}_{ring}}{\bar{e}_{alg}}$). The bold values indicate better performance in comparison with the ring algorithm}
\label{tab:alg-cmp}
\begin{tabular}{|p{1.7cm}|p{0.8cm}|p{0.8cm}|p{0.8cm}|p{0.8cm}|p{0.8cm}|p{0.8cm}|p{0.8cm}|p{0.8cm}|}\hline
Max delay$\rightarrow$ & 0 & 1 & 5 & 10 & 50 & 100 & 500 & 1000\\
Algorithm$\downarrow$ &  &  &  &  &  &  &  & \\\hline
Ring & 67.8 & 67.6 & 70.1 & 75.6 & 115.9 & 165.5 & 558.6 & 1047.0\\
\ & 1.00 & 1.00 & 1.00 & 1.00 & 1.00 & 1.00 & 1.00 & 1.00\\\hline
Rabenseifner & 76.3 & 75.9 & 82.9 & 85.7 & 125.0 & 181.6 & 572.0 & 1061.8\\
\ & 0.89 & 0.89 & 0.85 & 0.88 & 0.93 & 0.91 & 0.98 & 0.99\\\hline
SLT & 90.3 & 90.3 & 89.7 & 89.7 & \bf 101.0 & \bf 149.8 & \bf 542.1 & \bf 1031.9\\
\ & 0.75 & 0.75 & 0.78 & 0.84 & \bf 1.15 & \bf 1.10 & \bf 1.03 & \bf 1.01\\\hline
PRR & 70.9 & 70.8 & 71.1 & \bf 73.0 & \bf 101.0 & \bf 149.7 & \bf 541.7 & \bf 1031.5\\
\ & 0.96 & 0.95 & 0.99 & \bf 1.04 & \bf 1.15 & \bf 1.11 & \bf 1.03 & \bf 1.01\\\hline
\end{tabular}
\end{table}

Table~\ref{tab:prr-eth} presents the benchmark results related to PRR in comparison with the ring algorithm, executed on 48~nodes with 1\,Gbps Ethernet interconnecting network. While the absolute average elapsed time savings of PRR algorithm are higher for longer messages (up to 82\,ms for 8\,M), the high speedup values occur for all message sizes providing relative savings up to 15\%. We can observe that the lower delays cause lager speedup losses (up to 22\%, but for absolute time: 1.5\,ms only), which is especially visible for the smallest message size: 128\,K. In general, for the lower delays, the PRR is comparable with the ring (see Fig.~\ref{fig:bench-delay} for 1\,M of float numbers reduced data size).

\begin{table}[h]
\caption{Comparison of ring and PRR algorithms for 1\,Gbps Ethernet and 48~processes/nodes. Maximum delay is measured in ms and size in Kfloats (4$\times$KB). Each entry consists of two values: a difference of the average elapsed times in ms ($\bar{e}_{PRR}-\bar{e}_{ring}$) and speedup: a quotient of elapsed times ($\frac{\bar{e}_{ring}}{\bar{e}_{PRR}}$). The bold values indicate better performance in comparison with the ring algorithm}
\label{tab:prr-eth}
\begin{tabular}{|p{1.7cm}|p{0.8cm}|p{0.8cm}|p{0.8cm}|p{0.8cm}|p{0.8cm}|p{0.8cm}|p{0.8cm}|p{0.8cm}|}\hline
Max delay$\rightarrow$ & 0 & 1 & 5 & 10 & 50 & 100 & 500 & 1000\\\hline
Size$\downarrow$ &\multicolumn{8}{c|}{Only one process delayed}\\\hline
\ & 1.0 & 1.5 & \bf -5.0 & \bf -4.1 & \bf -4.3 & \bf -3.2 & \bf -0.9 & \bf -0.8\\
128 & 0.85 & 0.78 & \bf 1.53 & \bf 1.26 & \bf 1.08 & \bf 1.03 & \bf 1.00 & \bf 1.00\\\hline
\ & \bf -1.0 & \bf -0.3 & \bf -10.3 & \bf -12.6 & \bf -16.4 & \bf -11.6 & \bf -1.5 & \bf -1.7\\
512 & \bf 1.04 & \bf 1.01 & \bf 1.37 & \bf 1.41 & \bf 1.26 & \bf 1.10 & \bf 1.00 & \bf 1.00\\\hline
\ & 2.9 & 3.7 & 1.2 & \bf -3.2 & \bf -14.1 & \bf -15.2 & \bf -15.8 & \bf -16.0\\
1024 & 0.96 & 0.95 & 0.98 & \bf 1.04 & \bf 1.14 & \bf 1.10 & \bf 1.03 & \bf 1.02\\\hline
\ & 7.8 & 6.4 & 3.4 & \bf -1.7 & \bf -8.5 & \bf -23.9 & \bf -27.1 & \bf -27.7\\
2048 & 0.93 & 0.94 & 0.97 & \bf 1.02 & \bf 1.06 & \bf 1.13 & \bf 1.05 & \bf 1.03\\\hline
\ & 7.6 & 6.0 & 5.2 & 2.8 & \bf -20.9 & \bf -29.6 & \bf -38.1 & \bf -46.8\\
4096 & 0.96 & 0.97 & 0.97 & 0.99 & \bf 1.10 & \bf 1.12 & \bf 1.06 & \bf 1.04\\\hline
\ & 5.8 & 17.0 & 13.7 & 11.7 & \bf -29.0 & \bf -52.1 & \bf -75.7 & \bf -82.1\\
8192 & 0.98 & 0.95 & 0.96 & 0.97 & \bf 1.08 & \bf 1.14 & \bf 1.10 & \bf 1.07\\\hline
Size$\downarrow$ &\multicolumn{8}{c|}{All processes delayed randomly}\\\hline
\ & 1.2 & 1.0 & \bf -1.1 & \bf -1.1 & \bf -2.7 & \bf -1.0 & \bf -0.9 & \bf -1.3\\
128 & 0.82 & 0.85 & \bf 1.11 & \bf 1.09 & \bf 1.09 & \bf 1.02 & \bf 1.00 & \bf 1.00\\\hline
\ & 2.3 & \bf -0.3 & \bf -9.5 & \bf -9.3 & \bf -16.3 & \bf -11.5 & \bf -2.3 & \bf -2.9\\
512 & 0.91 & \bf 1.01 & \bf 1.37 & \bf 1.35 & \bf 1.38 & \bf 1.18 & \bf 1.01 & \bf 1.01\\\hline
\ & 2.9 & 3.8 & 1.7 & 0.7 & \bf -13.0 & \bf -13.2 & \bf -14.1 & \bf -14.5\\
1024 & 0.96 & 0.95 & 0.98 & 0.99 & \bf 1.17 & \bf 1.13 & \bf 1.05 & \bf 1.03\\\hline
\ & 7.1 & 8.3 & 4.8 & 3.0 & \bf -8.8 & \bf -21.2 & \bf -23.9 & \bf -24.5\\
2048 & 0.94 & 0.92 & 0.96 & 0.97 & \bf 1.08 & \bf 1.16 & \bf 1.07 & \bf 1.04\\\hline
\ & 6.1 & 0.1 & 3.2 & 5.6 & \bf -21.8 & \bf -19.0 & \bf -42.6 & \bf -39.8\\
4096 & 0.97 & 1.00 & 0.98 & 0.97 & \bf 1.11 & \bf 1.09 & \bf 1.11 & \bf 1.06\\\hline
\ & 13.8 & 16.3 & 6.8 & 7.1 & 0.4 & \bf -34.4 & \bf -75.9 & \bf -75.1\\
8192 & 0.96 & 0.95 & 0.98 & 0.98 & 1.00 & \bf 1.10 & \bf 1.15 & \bf 1.10\\\hline
\end{tabular}
\end{table}

The mode of the introduced PAT delay influences slightly the measured values. For larger delays and message sizes, when only one process is delayed the PRR algorithm provides slightly smaller relative savings than in the case when the delays were introduced for all processes, with the uniform probabilistic distribution, it is especially visible for 500--1000\,ms delays and message sizes of 2--8\,M of floats. However, in general, the PRR works fine for both modes of PAT delay.

\begin{figure}[!h]\sidecaption
\includegraphics[width=8cm]{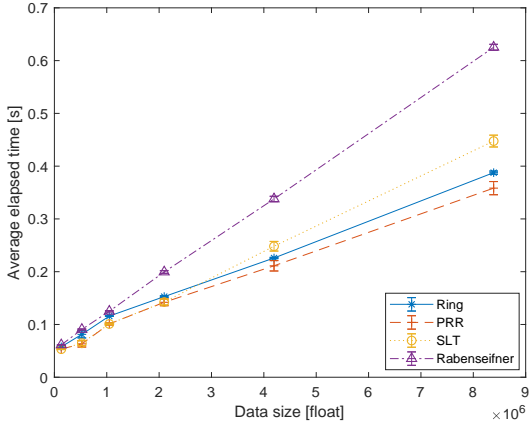} 
\caption{Benchmark results showing the algorithms' behavior related to the reduced data size. The experiments were performed on 48~nodes for 1\,Gbps Ethernet network, one process delayed, and the (maximum) delay equals 50\,ms. The error bars are set to $\pm{}2\sigma$ (95\% of the measurements for the normal distribution)}
\label{fig:bench-size}
\end{figure}

Fig.~\ref{fig:bench-size} presents the behavior of the tested algorithms in a context of the changing data size. While the PRR algorithm performs well for a wide range of size values, the SLT lags for the larger data, the threshold depends on the (maximum) delay, e.g. for 50\,ms the SLT is worse than the regular ring for 4+\,M of float numbers data size.

\begin{figure}[!h]\sidecaption
\includegraphics[width=8cm]{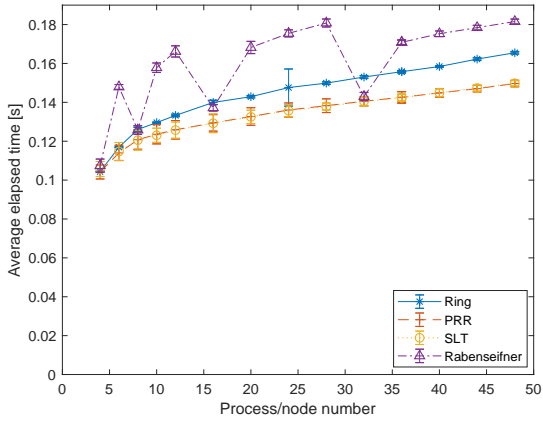} 
\caption{Benchmark scalability results for 1\,GB Ethernet network, one process delayed for 100\,ms, 1\,M of float numbers data size. The error bars are set to $\pm{}2\sigma$ (95\% of the measurements for the normal distribution)}
\label{fig:bench-proc}
\end{figure}

Fig.~\ref{fig:bench-proc} shows the tested algorithms performance related to the increase of the number of the nodes/processes exchanging reduced data. For 32 nodes, we observed interesting behavior of the Rabenseifner algorithm, which was designed for power of 2 node/process numbers. Both the PRR and SLT algorithms present stable speedup over the regular ring algorithm, proving good usability and high scalability for imbalanced PAPs.

\section{All-reduce PAP optimization for training of a deep neural network}\label{sec:ml}
In this section we present a practical application of the proposed method for a deep learning iterative procedure, implemented using tiny-dnn open-source library~\cite{tiny-dnn}. The example is focused on training of a convolutional neural network to classify graphical images (photos).

For the experiments we used a training dataset usually utilized for benchmarking purposes: CIFAR-10~\cite{cifar10}, it contains 60,000 32$\times$32 color images grouped in 10~classes. There are 50,000 training images, exactly 5,000 images per class. The other 10,000 test images are used to evaluate training results. In our example we assess the performance of the distributed processing, especially collective communication, thus we did not need to use the test set.

We test the network architecture 8~layers: 3~convolutional, 3~average pooling, and 2~fully connected ones, the whole model has 145,578~parameters of float size. Each process trains such a network and after each training iteration it is averaged over the other processes, a similar procedure was described in~\cite{Dean2012}, with the distinction in using a separated parameter server. The mini-batch size was set to 8~images per node, what gives 390~iterations in total.

The training program was implemented in C++ language (v. C++11) using tiny-dnn library~\cite{tiny-dnn}. It was chosen because it is very lightweight: header only, easy to install, dependency-free and has an open-source license: BSD 3-Clause. The above features made it easy to introduce the required modification: a callback function for each neural network layer, called during the distributed SGD (Stochastic Gradient Descent~\cite{Dean2012}) training. The function is used for progress monitoring, where the egde() function (see section~\ref{sec:model}) is called just before the last layer is processed, which takes about 44\% of computation time of the iteration. Additionally, the computation part of each iteration is performed in parallel, using POSIX threads~\cite{pthreads} executed on available (24) cores (provided by two processors with Hyper Threading switched off).

We tested 3~all-reduce algorithms: ring, sorted linear tree (SLT) and pre-reduced ring (PRR). The PAP framework, including estimation of computation time and warm-up was initiated only for the latter two. The benchmark was executed 128~times for each algorithm, and each execution consists of 390~training iterations with all-reduce function calls. The tests were executed in an HPC cluster environment, consisting of 16~nodes with 1\,Gbps Ethernet interconnecting network (the configuration is described in section~\ref{sec:bench-setup}). 

The results of the benchmark are presented in Table~\ref{tab:cifar10}. The PRR has the lowest time of communication: 31.9\,ms (average elapsed time of all-reduce) as well as the total time of training: 78.6\,s, SLT is slightly worse having 33.1\,ms and 78.8\,s respectively. These algorithms were compared to a typical ring~\cite{Thakur2005} implementation, where the average elapsed time equals 35.7\,ms and the total training took 81.9\,s. In context of the average elapsed time, the PRR and SLT algorithms are faster for 12.1\% and 7.9\%, while for the training total time 4.2\% and 4.0\% respectively.

\begin{table}[h]
\caption{Times and speedup of Cifar10~\cite{cifar10} benchmark execution. The times are presented in ms, and the speedup is calculated in comparison with a ring algorithm implementation}
\label{tab:cifar10}
\begin{tabular}{|p{1.4cm}|p{2.1cm}|p{2.1cm}|p{2.1cm}|p{2.1cm}|}
\hline
Algorithm & All-reduce average elapsed time & All-reduce speedup & Training total time & Training speedup\\\hline
Ring & 35.7 & 1.000 & 81,900 & 1.000\\\hline
SLT & 33.1 & 1.079 & 78,768 & 1.040\\\hline
PRR & 31.9 & 1.121 & 78,604 & 1.042\\\hline
\end{tabular}
\end{table}

The above results seem to be just a slight improvement, however in the current, massive processing systems, e.g. neural networks, which are trained using thousands of compute nodes, consuming MegaWatts of energy, with budgets of millions of dollars, introducing 4\% computing time reduction, without additional resource demand can provide great cost savings.

\section{Final remarks}
The proposed algorithms provide optimizations for all-reduce operations executed in environment of imbalanced PAPs. The experimental results show improved performance and good scalability of the proposed solution over currently used algorithms. The real case example (machine learning using distributed SGD~\cite{Dean2012}) shows usability of the prototype implementation with sum as the reducing operation. The solution can be used in a wide spectrum of applications using iterative computation model, including many machine learning algorithms.

The future works cover the following topics:
\begin{itemize}
\item evaluation of the method for a wider range of interconnecting network speeds and larger number of nodes using a simulation tool e.g.~\cite{Czarnul2017,Proficz2016},
\item expansion of the method for other collective communication algorithms, e.g. all-gather,
\item a framework for automatic PAP detection and proper algorithm selection, e.g. providing a regular ring for balanced PAPs and PRR for imbalanced ones,
\item introduction of the presented PAT estimation method for other purposes e.g. asynchronous SDG training~\cite{Dean2012} or deadlock and race detection in distributed programs~\cite{Krawczyk2000},
\item deployment of the solution in a production environment.
\end{itemize}

We believe that the ubiquity of the imbalanced PAPs in HPC environments~\cite{Faraj2008} will cause a fast development of new solutions related to this subject.

\bibliographystyle{plain}
\bibliography{paper}

\end{document}